\documentclass[final]{siamltex}
\usepackage{amsmath,amssymb,epsfig}

\newcommand{\bv}{\boldsymbol{\beta}}
\newcommand{\q}{{\tt q}}

\begin{document}
\title{Numerical Solution of the Small Dispersion Limit of the
Camassa-Holm and Whitham Equations and Multiscale Expansions}

\author{S.~Abenda\thanks{Dipartimento di Matematica e CIRAM,
                         Universit\`a di Bologna,
                         Via Saragozza 8, I-40123 Bologna BO, Italia}
                         \and
T.Grava\thanks{SISSA, Via Beirut 2-4, I-34100 Trieste, Italy} \and
C.~Klein\thanks{Institut de Math\'ematiques de Bourgogne,
                Universit\'e de Bourgogne, 9 avenue Alain Savary, 21078 Dijon
                Cedex, France}
    {\tt christian.klein@u-bourgogne.fr}}

\maketitle

\begin{abstract}
   The small dispersion limit of solutions to the Camassa-Holm (CH)
   equation is characterized by the appearance of a zone of rapid
   modulated oscillations. An asymptotic description of these
   oscillations is given, for short times, by the one-phase
   solution to the CH equation,
   where the branch points of the corresponding elliptic curve depend on the physical
   coordinates via the  Whitham equations. We present a conjecture for
   the phase of the asymptotic solution. A numerical
   study of this limit for smooth hump-like initial data provides strong
   evidence for the validity of this conjecture. We present a
   quantitative numerical comparison between the CH
   and the asymptotic solution. The dependence on the small dispersion
   parameter $\epsilon$ is studied in the interior and at the
   boundaries of the Whitham zone. In the interior of the zone, the
   difference between CH and asymptotic solution is of the order
   $\epsilon$, at the trailing edge of the order $\sqrt{\epsilon}$
   and at the leading edge of the order $\epsilon^{1/3}$. For the
   latter we present a multiscale expansion which describes the
   amplitude of the oscillations in terms of the Hastings-McLeod
   solution of the Painlev\'e II equation. We show numerically that
   this multiscale solution provides an enhanced asymptotic description near the
   leading edge.
\end{abstract}



\pagestyle{myheadings} \thispagestyle{plain}

\section{Introduction}
The Camassa-Holm (CH) equation
\begin{equation}
    u_{t}+(3u+2\nu)u_{x}-\epsilon^{2}(u_{xxt}+2u_{x}u_{xx}+uu_{xxx})=0,
    \quad x\in\mathbb{R}, \quad t>0,
    \label{CH}
\end{equation}
was discovered by Camassa and Holm  \cite{CH} as a model for
unidirectional propagation of waves in shallow water, $u(x,t)$
representing the height of the free surface about a flat bottom,
$\nu$  being a constant related to the critical shallow water
speed and $\epsilon $ a constant proportional to the mean water
depth \cite{DGH}. Equation (\ref{CH}) had been previously found by
Fokas and Fuchssteiner \cite{FokFuch} using the method of
recursion operators and shown to be a bi-hamiltonian equation with
an infinite number of conserved functionals. It was also
rediscovered by Dai \cite{Dai} as a model for nonlinear waves in
cylindrical hyperelastic rods, with $u(x,t)$ representing the
radial stretch relative to a pre-stressed state. Equation
(\ref{CH}) finally also arises in the study of the motion of a
non-Newtonian fluid of second grade in the limit when the
viscosity tends to zero \cite{Busu}. A class of two--component
generalizations of the CH equation has been recently obtained in
\cite{Falqui}.

The initial value problem for (\ref{CH})
\[
u(x,0)= u_0(x),\quad\quad x\in \mathbb R,
\]
presents interesting features: first for $\nu=0$ there may exist
peakons, i.e., non--smooth solutions, second even for a smooth
initial datum $u_0(x)$ the wave-breaking phenomenon may occur,
that is the solution $u(x,t)$ remains bounded while its slope
becomes unbounded in finite time. This phenomenon was first
noticed for $\nu=0$ by Camassa and Holm \cite{CH} who showed that
for smooth and odd initial datum $u_0(x)$ such that $u_0(x)>0$ for
$x<0$ and $u_0^{\prime}(0)<0$, the slope $ u_x(x,t)$ is driven to
$-\infty$ in finite time. In the case $\nu=0$, under the
hypothesis that
\[
m_0(x):=u_0(x)-\epsilon^2
\partial_{xx} u_0(x)\] is smooth and summable,
McKean \cite{Kean1} proves that the wave-breaking phenomenon
occurs if and only if some portion of the positive part of
$m_0(x)$ lies to the left of some portion of the negative part of
$m_0(x)$. The wave breaking phenomenon occurs also for $\nu \not
=0$ and, in particular Constantin \cite{Const} shows that for
initial data $u_0(x)$ in the Sobolev space $H^s (\mathbb R)$, $s
\ge 3$ there exists a unique solution to (\ref{CH}) $u\in {\cal
C}^0 ([0,T[, H^s (\mathbb R)) \cap{\cal C}^1 ([0,T[, H^{s-1}
(\mathbb R))$ defined for some maximal time $T>0$; moreover
$T<\infty$ if and only if $\liminf_{t\to T}  \{ \min_{x\in \mathbb
R} u_x(x,t) \}=-\infty$, i.e., for $\nu \not =0$, singularities in
the solution may arise only in the form of wave breaking.

\smallskip

In this manuscript we are interested in studying the behaviour of
the solution $u(x,t,\epsilon)$ of the Cauchy problem of the CH
equation as $\epsilon\rightarrow 0$ for smooth initial data for
which breaking does not occur. To this end, we suppose that
$u_0(x)$  is in the Schwartz class, has a single negative hump and
satisfies the following non--breaking condition
\begin{equation}\label{nonbreak}
m_0(x)+\nu> 0,\quad x\in\mathbb{R}.
\end{equation}
In this case for any $t>0$, $m(x,t;\epsilon)=u(x,t)-\epsilon^2
\partial_{xx} u(x,t)$ is in the Schwartz class and
$m(x,t;\epsilon)+\nu>0$  \cite{ConstLen}.

\begin{figure}[!htb]
\centering \epsfig{figure=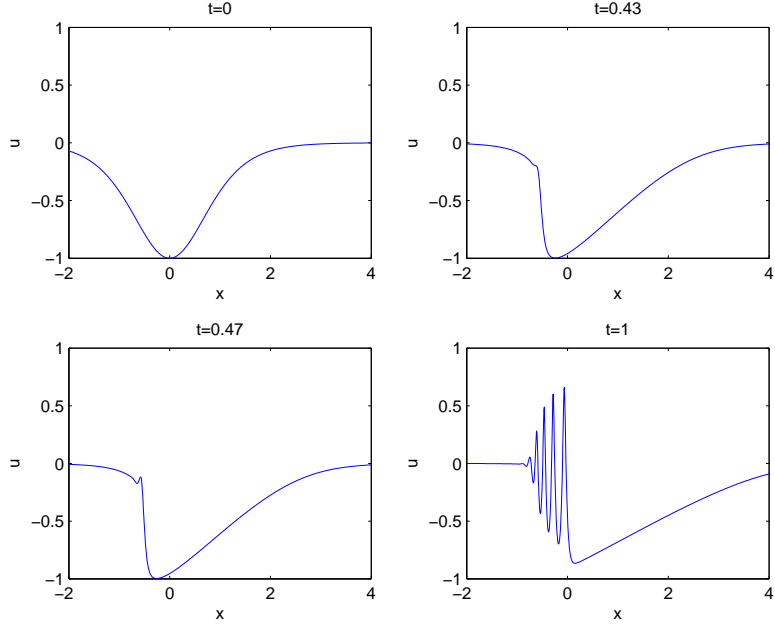, width=12cm}
\caption{Solution to the CH equation for
$u_{0}=-\mbox{sech}^{2}x$,  $\nu=1.2$ and $\epsilon=10^{-1.5}$ for
several values of $t$.} \label{chtime}
\end{figure}

For initial data in this class and $\epsilon \ll 1$,  Grava and
Klein \cite{grakle12} show that the numerical solution
$u(x,t,\epsilon)$ of (\ref{CH}) develops a zone of fast
oscillations, as  for the small dispersion limit of the
Korteweg-de Vries (KdV) equation, see Fig.~\ref{chtime}.

In this paper, following the works  of Gurevich and Pitaevskii
\cite{GP}, Lax and Levermore \cite{lale}, Deift, Venakides and
Zhou \cite{devezo}, we claim and give numerical support for our
claim that the description of the small dispersion limit of the CH
equation follows 1)-3) below:

\medskip

1) For $0\le t <t_c$, where $t_c$ is a  critical time,  the
solution $u(x,t,\epsilon)$ of the CH Cauchy problem is
approximated as  $\epsilon\rightarrow 0 $, by $u(x,t)$ which solves the
Hopf equation
\begin{equation}\label{Hopf}
u_t + (3u+2\nu) u_x =0.
\end{equation}
Here $t_c$ is the time when the first point of  gradient
catastrophe appears in the solution to the Hopf equation
\[
u(x,t) = u_0 (\xi), \quad\quad x = (3u_0 (\xi) +2\nu) t +\xi,
\]
and it is given by the relation
\[
t_c = \frac{1}{\max_{\xi\in {\mathbb R}} [-3u^{\prime}_0(\xi)]}.
\]

\medskip

2) There exists  $T>t_c$ such that for $t_c<t<T$, the solution of
the CH equation is characterized by the appearance of an interval
of rapid oscillations. If $f'''_-(u)<0$ where $f_-$ is the inverse
of the decreasing part of the initial datum $u_0(x)$ and $u_c$ is
the critical point,  these oscillations are described in the
following way. As $\epsilon\rightarrow 0$, the interval of the
oscillatory zone is given by $[x^{-} (t), x^+(t)]$ where
$x^{\pm}(t)$ are determined  from the initial datum and satisfy
the condition $x^-(t_c)=x^+(t_c)=x_c$, with $x_c$ the coordinate
of the point of the gradient catastrophe of the Hopf solution.  In the (x,t) plane $x^\pm(t)$ describe a cusp-shape region.
Outside the interval $[x^{-} (t), x^+(t)]$ the leading order
asymptotics is given by the solution of the Hopf equation. Inside
the interval $(x^{-} (t), x^+(t))$ we claim that the solution
$u(x,t,\epsilon)$ is approximately described, for small $\epsilon$
by the one--phase solution of CH which may be expressed in
implicit form in terms of elliptic functions as
\begin{equation}
\label{abe:ellip}
 \left\{
\begin{split}
\displaystyle u(x,t,\epsilon) &\simeq \beta_1+\beta_2+\beta_3+2\nu
-2(\beta_3+\nu)\frac{\Lambda(s,\rho)}{K(s)}+\frac{2}{I_0}\frac{d}{dz}\log
\frac{\vartheta_3(z-p,\tau)}{\vartheta_3(z+p,\tau)},\\
\displaystyle \zeta &=\frac{k}{\epsilon}\left(
x-(\beta_1+\beta_2+\beta_3+2\nu)t-q\right) = 2\pi
z-k\log\frac{\vartheta_3(z-p,\tau)}{\vartheta_3(z+p,\tau)},
\end{split}
\right. \end{equation} with $\beta_1>\beta_2>\beta_3>-\nu$, $K(s)$
and $\Lambda(s,\rho)$  the complete Jacobi elliptic integrals of
the first and third kind of modulus $s$, respectively,  where
\begin{equation}
\label{abe:modulus} s^2 =
\frac{(\beta_1+\nu)(\beta_2-\beta_3)}{(\beta_2+\nu)(\beta_1-\beta_3)},
\quad\quad\rho =\frac{\beta_2-\beta_3}{\beta_2+\nu}.
\end{equation}
The quantities $I_0$,  $p$ and the wave-number $k$ are Abelian
integrals defined respectively in (\ref{abe:kappa}),
(\ref{abe:p0}) and (\ref{abe:semi}), the phase shift $q(\bv)$  is
defined in (\ref{q}). $\vartheta_3(z,\tau)$ is the third Jacobi
theta function of modulus $\tau =iK(\sqrt{1-s^2})/K(s)$ (see
\cite{WW} and  (\ref{vartheta3})).

\medskip

For constant values of the $\beta_i$, (\ref{abe:ellip}) is an
exact solution to CH in implicit form (see \cite{agk} and section
\ref{11}). Moreover, for $-\nu<\beta_3<\beta_2<\beta_1$, such
solution is real--periodic and analytic in $\zeta$, so it is
appropriate to call it the one--phase solution. However, unlike
the KdV case, it cannot be extended to meromorphic function on the
$\zeta$ complex plane. Indeed, the r.h.s of the first equation in
(\ref{abe:ellip}) is an elliptic function in $z$, while
$\zeta=\zeta(z)$ is real analytic and invertible for real $z$, but
not meromorphic in $\mathbb C$.

\medskip

In the description of the leading order asymptotics of
$u(x,t,\epsilon)$, as $\epsilon\to 0$, the quantities $\beta_i$
depend on $x,t$ and evolve according to the CH modulation
equations which were derived in \cite{abegra} for the one--phase
periodic solution. In terms of the Riemann invariants
$\beta_3<\beta_2<\beta_1$ they take the form
\begin{equation}\label{abe:whi}
{\partial_t} \beta_i + C_i (\bv) {\partial_x} \beta_i =0,\quad\quad
C_i(\bv) =\frac{\partial_{\beta_i} \omega (\bv)}{\partial_{\beta_i} k
(\bv)}, \;\; i=1,2,3,
\end{equation}
with  $\bv =(\beta_1,\beta_2,\beta_3)$ and
$\omega=(\beta_1+\beta_2+\beta_3+2\nu)k$. Unlike the KdV case, the
Whitham equations for CH are not strictly hyperbolic and this fact
gives some technical difficulties in their numerical and
analytical treatment.
\smallskip

3) Near the left boundary of the cusp-shape region,  in the double--scaling limit $x\rightarrow x^-(t)$ and
$\epsilon\rightarrow 0$, in such a way that
\[
(x-x^-(t))\epsilon^{-2/3}
\]
remains finite, the asymptotic solution of the CH equation is
given by
\[
u(x,t,\epsilon)=u(x^-(t),t)+
\epsilon^{\frac{1}{3}}a(x,t)\cos(\psi(x,t,\epsilon)/\epsilon)
+O(\epsilon^{\frac{2}{3}}),
\]
where $u(x,t)$ is the solution of the Hopf equation (\ref{Hopf}),
the phase $\psi$ is given in (\ref{psi}) and the function
$a(x,t,\epsilon)$ is, up to shifts and rescalings, the
Hastings-Mcleod solution to the Painlev\'e-II equation \cite{HMcL}
\[
A_{zz}=zA+2A^3,\quad A, z\in\mathbb{C},
\]
determined uniquely by the boundary conditions
\[
A(z) \approx Ai(z) \quad (z\to +\infty),\quad\quad
A(z)\approx\sqrt{-z/2} \quad (z\to -\infty)
\]
with $Ai(z)$ the Airy function. Such a solution is real and pole
free for real values of $z$.

\begin{figure}[!htb]
\centering \epsfig{figure=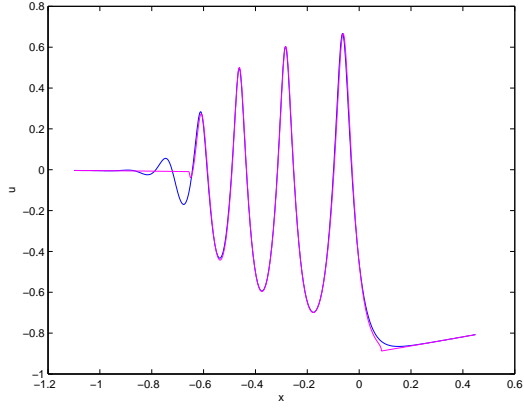, width=8cm}
\caption{Solution to the CH equation for the initial datum
$u_{0}=-\mbox{sech}^{2}x$ and $\nu=1.2$ at $t=1$ in blue and the
corresponding asymptotic solution via the Hopf equation and the
one-phase CH solution solution.} \label{ch1e3asym}
\end{figure}

We verify numerically the validity of the above asymptotic expansions
1), 2)
and 3) for the initial datum $u_0(x)=-1/\cosh^2x$ and different
values of $\nu$, see for instance Fig.~\ref{ch1e3asym}.\\
{\bf Remark.}
The asymptotic description of $u(x,t,\epsilon)$ as
$\epsilon\rightarrow 0$  near the critical time $t=t_c$ has been
conjectured in \cite{Du} and studied numerically in \cite{grakle}. The asymptotic description of $u(x,t,\epsilon)$ as
$\epsilon\rightarrow 0$  at the right boundary of the cusp-shape region, namely near $x^+(t)$ has not yet been studied even for the KdV case.

The paper is organized as follows: In the next section we compute
the one--phase solution to CH, in section \ref{1} we obtain the
small amplitude limit of the one--phase solution to the CH
equation. In section \ref{2} we perform a multiple scale analysis
of the CH equations near the leading edge $x^-(t)$ of the
oscillatory zone. In section \ref{3} we present a numerical
comparison of the small dispersion limit  of the  CH solution with
the asymptotic formula (\ref{abe:ellip}). A quantitative numerical
comparison of the CH solution and the multiscale solution is
presented in section \ref{4}.

\section{The one-phase  solution to the Camassa--Holm equation}\label{11}
Let us look for a one--phase real--periodic travelling wave
solution to (\ref{CH}) of the form
\[
u(x,t)= (c-2\nu) -2\eta(\zeta), \quad\quad \zeta =\frac{kx-\omega
t +\phi_0}{\epsilon},\quad\quad c=\frac{\omega}{k}
\]
where $k$ is the wave number, $\omega$ the frequency and $\phi_0$
is a phase to be determined from the initial conditions. When we
plug $\eta(\zeta)$ into the CH equation (\ref{CH}), after
integration we get
\begin{equation}
\label{periodic2}   k^2(\eta+\nu) \eta_{\zeta}^2- \eta^3
+(c-2\nu)\eta^2 -2B\eta+2A=0,
\end{equation}
where $A$ and $B$ are constants of integration.
The CH one-phase solution $u(x,t)=(c-2\nu) -2\eta(\zeta)$
is then obtained by inverting the differential of the  third kind
\begin{equation}
\label{abe:solint} \zeta=k
\int_{\eta_0}^\eta\displaystyle\frac{(\xi+\nu) d\xi}{
\sqrt{((\xi+\nu)(\xi^3+ (2\nu-c)\xi^2 +2B\xi-2A)}}=
k\int_{\eta_0}^{\eta} \frac{(\xi +\nu)d\xi}{\sqrt{(\xi+\nu)
\prod_{i=1}^3 (\xi -\beta_i)}},
\end{equation}
where $ \beta_1+\beta_2+\beta_3 =c-2\nu$. Let
$-\nu<\beta_3<\beta_2<\beta_1$, so that $\eta(\zeta)$ is real
periodic in the interval $[\beta_3,\beta_2]$. Since $\displaystyle
\frac{d\zeta}{d\eta}$ has constant sign for
$\eta\in[\beta_3,\beta_2]$, by a standard argument, the (real)
inverse function $\eta(\zeta)$ exists and is monotone in $\zeta\in
[0,Z]$, where $Z$ is the half period of the travelling wave
solution. To invert (\ref{abe:solint}) and obtain
$\eta=\eta(\zeta)$ let us introduce the elliptic curve
\begin{equation}\label{ellicurv}
{\mathcal E}:\; \{ w^2=R(\xi)=(\xi+\nu)\prod_{i=1}^3 (\xi-\beta_i)
\},\end{equation} with  homology basis of cycles $a,b$ as in the
figure below and let
\begin{figure}[!htb]
\centering
\epsfig{figure=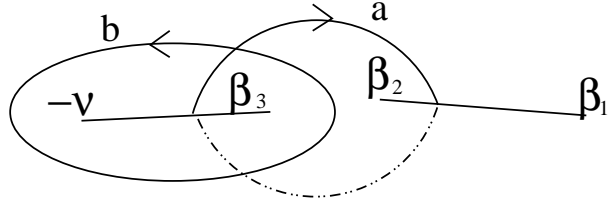, width=8cm}
\caption{The homology basis} \label{homology}
\end{figure}
\begin{equation}\label{abe:kappa}
\displaystyle
{\phi} = \frac{d\xi}{I_0 w},\quad
I_0= \oint_{a} \frac{d\xi}{w} =
\frac{4K(s)}{\sqrt{(\beta_2+\nu)(\beta_1-\beta_3)}},
\end{equation}
be the normalized holomorphic differential on ${\mathcal E}$, with
$K(s)$ the complete Jacobi elliptic integral of modulus $s$
\begin{equation}
\label{abe:modulus1} s^2 =
\frac{(\beta_1+\nu)(\beta_2-\beta_3)}{(\beta_2+\nu)(\beta_1-\beta_3)}.
\end{equation}
Then inversion of the normalized holomorphic differential
\[z= \int_{-\nu}^\eta\phi
\]
is given by the Jacobi theorem \cite{Du0} and takes the form
\begin{equation}
\label{abe:jacobi} \eta(z)=\oint_a \xi \phi
-\dfrac{1}{I_0}\dfrac{d}{dz}
\log\dfrac{\vartheta_3(z-p_0+\frac{\tau}{2}+\frac{1}{2})}
{\vartheta_3(z+p_0+\frac{\tau}{2}+\frac{1}{2})},
\end{equation}
where
\[
\oint_a \xi \phi = -\nu+(\nu+\beta_3)\frac{\Lambda(s,\rho)}{K(s)},
\]
\begin{equation}\label{elli3}\Lambda(s,\rho) = \int_0^1 \frac{dz}{(1-\rho
z^2)\sqrt{(1-z^2)(1-s^2z^2)}},\quad\quad
\rho=\frac{\beta_2-\beta_3}{\beta_2+\nu}, \end{equation}
$\vartheta_3(z)$ is the third
Jacobi-theta function defined by the Fourier series
\begin{equation}\label{vartheta3}
\vartheta_3(z,\tau)=\sum_{n=-\infty}^{+\infty} \exp \left[i\pi n^2
\tau +2\pi i nz \right],\quad\quad \tau =i
\frac{K^{\prime}(s)}{K(s)},
\end{equation}
$K^{\prime} (s) = K(\sqrt{1-s^2})$ and $p_0$ is defined by
\begin{equation}\label{abe:p0}
p_0=\int_{-\nu}^{\infty^+}\phi= \frac{1}{2} +p, \quad\quad
p=\int_{\beta_1}^{\infty^+}\phi = \frac{1}{2K(s)}
\int_{0}^{\sqrt{\frac{\beta_2+\nu}{\beta_1+\nu}}}\frac{dz}{\sqrt{(1-z^2)(1-s^2
z^2)}}.
\end{equation}
The function $\eta(z)$ is an elliptic function in $z$ with periods $1$
and $\displaystyle \tau$
and it  is real periodic in $z$ along the
complex line $z= \tau/2 + z^{\prime}$, $z^{\prime} \in
\mathbb{R}$.

Finally, inserting (\ref{abe:jacobi}) into
(\ref{abe:solint}), we get
\begin{equation}
\label{solution} \displaystyle \zeta =k
I_0\int^{z+\tau/2}_{\tau/2} (\eta+\nu)dz= kI_0
(\nu+\beta_3)\frac{\Lambda(s,\rho)}{K(s)}  z-
k\log\frac{\vartheta_{3}(z-p,\tau)}{\vartheta_{3}(z+p,\tau)}.
\end{equation}
Equations (\ref{abe:jacobi}) and (\ref{solution}) give the
one--phase solution $u(\zeta)$ in implicit form. The real half
period, $Z$, of $u(\zeta)$ is
\begin{equation}\label{abe:semi}
Z= k I_0\int^{(\tau+1)/2}_{\tau/2} (\eta(z)+\nu)dz.
\end{equation}
Normalizing the half period $Z$ to $\pi$, from (\ref{abe:semi}) we
obtain the wave number and the frequency
\begin{equation}
\begin{split}
\label{komega}
k&=2\pi\left(\oint_a\displaystyle\frac{(\xi+\nu)d\lambda}{\sqrt{R(\xi)}}
\right)^{-1}=\frac{\pi\sqrt{(\beta_2+\nu)(\beta_1-\beta_3)}}{2(\beta_3+\nu)
\Lambda(s,\rho)} ,
\\
\omega&=(2\nu+\beta_1+\beta_2+\beta_3)k.
\end{split}
\end{equation}
Summarizing, the one--phase real nonsingular solution to the CH
equation takes the form
\begin{equation}
\label{abe:ellisol}
\left\{
\begin{array}{l}
\displaystyle u(x,t,\epsilon) = \beta_1+\beta_2+\beta_3+2\nu
-2(\beta_3+\nu)\frac{\Lambda(s,\rho)}{K(s)}+\frac{2}{I_0}\frac{d}{dz}\log
\frac{\vartheta_3(z-p,\tau)}{\vartheta_3(z+p,\tau)},\\
\displaystyle \frac{1}{\epsilon}\left(
x-(\beta_1+\beta_2+\beta_3+2\nu)t-q\right) = \frac{2\pi
z}{k}-\log\frac{\vartheta_3(z-p,\tau)}{\vartheta_3(z+p,\tau)}.
\end{array}
\right.
\end{equation}
Observe from the above formula that $u(x,t,\epsilon)$ is not a
meromorphic function of $x$ and $t$. An alternative method to
obtain $u(x,t,\epsilon)$ is discussed in \cite{AlFe} where the
solution is expressed in terms of conveniently generalized
theta-functions in two variables, which are constrained to the
generalized theta-divisor of a 2--dimensional generalized
Jacobian.

The envelope of the oscillations is obtained by the maximum and
minimum values of the theta function and gives
\[
\begin{split}
u_{max} &= \beta_1+\beta_2+\beta_3 +2\nu
-2(\beta_3+\nu)\frac{\Lambda(s,\rho)}{K(s)}+\frac{4}{I_0}
\frac{\vartheta^{\prime}_3(p+\frac{1}{2})}{\vartheta_3(p+\frac{1}{2})},
\\
u_{min}  &= \beta_1+\beta_2+\beta_3 +2\nu
-2(\beta_3+\nu)\frac{\Lambda(s,\rho)}{K(s)}+\frac{4}{I_0}
\frac{\vartheta^{\prime}_3(p)}{\vartheta_3(p)}.
\end{split}
\]

\section{Small amplitude limit of the one-phase  solution\label{1}}

We analyze the one-phase solution (\ref{abe:ellisol}) near the
leading edge $x^-(t)$, namely when the oscillations go to zero.
For this purpose we need to study the CH modulation equations more
in detail.
\subsection{Camassa-Holm modulation equations}
In \cite{abegra}, Abenda and Grava constructed the one-phase CH
modulation equations and showed that $\beta_i$, $i=1,2,3$ are the
Riemann invariants. The CH modulation equations take the Riemann
invariant form
\begin{equation}\label{whithamu}
\partial_t \beta_i + C_i({\bv}) \partial_x \beta_i=0, \quad\quad i=1,\dots,3,
\end{equation}
where the speeds $C_i(\bv)$ in  (\ref{abe:whi})
are explicitly given by the formula \cite{abegra}
\begin{equation}
\label{Ci1}
\begin{split}
&C_1(\bv)=\beta_1+\beta_2+\beta_3+2\nu+2\displaystyle
\frac{(\beta_3+\nu)(\beta_1-\beta_2)
\Lambda(s,\rho)}{(\beta_2+\nu)E(s)},
\\
&C_2(\bv)=\beta_1+\beta_2+\beta_3+2\nu+\displaystyle
\frac{2(\beta_2-\beta_3)\Lambda(s,\rho)}{ K(s)-\displaystyle
\frac{(\beta_2+\nu)(\beta_1-\beta_3)}{(\beta_3+\nu)(\beta_1-\beta_2)}E(s)}
\\
&C_3(\bv)=\beta_1+\beta_2+\beta_3+2\nu+2\displaystyle
\frac{(\beta_3+\nu)(\beta_3-\beta_2) \Lambda(s,\rho)}
{(\beta_2+\nu)[K(s)-E(s)]},
\end{split}
\end{equation}
with $\bv =(\beta_1,\beta_2,\beta_3)$, $K(s)$, $\Lambda(s,\rho)$
as before and $E(s)$ the complete elliptic integral of the second
kind with modulus $s^2$.
In the limit when  two Riemann  invariants coalesce, the  modulation
equations reduce to the Hopf equation
\[
\partial_t u + (3u+2\nu)\partial_x u=0.
\]
The CH modulation equations are integrable via the generalized hodograph
transform introduced by Tsarev \cite{T} ,
\begin{equation}
\label{abe:hodosol} x=-C_{i}(\bv)\,t+w_{i}(\bv) \,\quad i=1,2,3\,,
\end{equation}
which gives the solution of (\ref{whithamu}) in implicit form. The
formula of the $ w_i(\bv) $ for the specific CH case, has been
obtained in
 \cite{abegra}, \cite{GPT}:
  \begin{equation}\label{abe:wi}
w_i (\bv)= q(\bv) + \left( C_i(\bv)
-(\beta_1+\beta_2+\beta_3+2\nu)\right) \partial_{\beta_i} q(\bv),
\quad\quad i=1,2,3,
\end{equation}
where the function $q=q(\bv)$ is given by
\begin{equation}\label{q}
q(\beta_1,\beta_2,\beta_3) = \frac{1}{2\sqrt{2}}\int_{-1}^1 \int_{-1}^1 d \mu
d\lambda \frac{  f_-\left( \frac{1+\mu}{2} \left(
\frac{1+\lambda}{2} \beta_3 +\frac{1-\lambda}{2}\beta_2 \right)
+\frac{1-\mu}{2} \beta_1\right)}{\sqrt{1-\mu}\sqrt{1-\lambda^2}},
\end{equation}
with $f_-$ the inverse of the decreasing part of the initial datum
$u_0(x)$. The above formula of $q$ is valid for  $x<x_M(t)$  where
$x_M(t)$ is the coordinate of the minimum value of the solution
$u(x,t)$ of the Hopf equation. For $x>x_M(t)$ the corresponding
formula for $q$ contains also the increasing part of the initial
datum  (see \cite{feiran}).
In \cite{GPT}  it is shown that if $f_-'''(u_c)<0$, then the solution $\beta_1(t)>\beta_2(t)>\beta_3(t)$ of the system (\ref{abe:wi}) exists for some time $t>t_c$.
\smallskip

In order  to take the small amplitude limit of the solution
(\ref{abe:ellip}), we rewrite the system (\ref{abe:hodosol}) in
the equivalent form \cite{GPT}
\begin{equation}\label{abe:hodonodeg}
\left\{ \begin{array}{l} (C_1 t + w_1 -x )\alpha =0,\\
\\
C_2 t+ w_2 -x=0,\\
\\
\displaystyle \frac{(C_2-C_3) t + (w_2-w_3)}{\beta_2-\beta_3} =0,
\end{array}\right.
\end{equation}
where
\begin{equation}\label{alpha}
\alpha =
\frac{(\beta_2+\nu)E(s)}{(\beta_1-\beta_2)(\beta_3+\nu)\Lambda(\rho,s)}\left(
\beta_1-\frac{\beta_2+\beta_3}{2}\right),
\end{equation}
with $s$ and $\rho$ as in (\ref{abe:modulus}) and (\ref{elli3}),
and we perform the limit $\delta\to 0$, where
\[
\beta_2 = v+\delta, \quad\quad \beta_3 = v-\delta,\quad\quad \beta_1=u.
\]
Let
\begin{equation}\label{abe:quv}
\begin{split}
Q(u,v) &= q(u,v,v)= \frac{\sqrt{2}}{4} \int_{-1}^1
\frac{d\mu}{\sqrt{1-\mu}} f_-\left( \frac{1+\mu}{2}v +
\frac{1-\mu}{2}u\right),\\
\Phi(u,v) &= \partial_u Q(u,v)+\partial_v Q(u,v)
=\frac{\sqrt{2}}{4} \int_{-1}^1 \frac{d\mu}{\sqrt{1-\mu}}
f_-^{\prime}\left( \frac{1+\mu}{2}v + \frac{1-\mu}{2}u\right),
\end{split}
\end{equation}
then the following identities hold
\begin{equation}\label{ident}
\begin{split}
f_-(u)&= Q(u,v)+ 2(u-v)\partial_u Q(u,v),\\ \partial_{uv}
Q(u,v)&= \frac{2\partial_u Q(u,v)-\partial_vQ(u,v)}{2(u-v)},\\
\partial_{uvv} Q(u,v)
&=\frac{4\partial_{uv}Q(u,v)-\partial_{vv}Q(u,v)}{2(u-v)},
\end{split}
\end{equation}
Substituting (\ref{abe:quv}), (\ref{ident}) and the expansion of
the elliptic integrals $E(s)$, $K(s)$ and $\Lambda(s,\rho)$ as
$s,\rho\rightarrow 0$ \cite{GPT}
into (\ref{abe:hodonodeg}), we arrive to the system
\begin{equation}\label{abe:tay1}
\begin{split}
 x =& (3u+2\nu)t +
f_-(u)-\frac{\delta^2}{2}\frac{(3t+\Phi(u,v))(2v+u+3\nu)}{(u-v)^2}
+O(\delta^3),\\
\displaystyle x =&(3u+2\nu)t +f_-(u) -2\frac{u-v}{u+\nu}\left( t
(u+2v+3\nu) +\Phi(u,v)(v+\nu) +\partial_uQ(u,v)
(u-v) \right)\\
&+\delta\left(
\frac{(2v-u+\nu)(3t+\Phi(u,v))-2(u-v)(v+\nu)\partial_v
\Phi(u,v)}{2(u+\nu)}\right)
-\delta^2\left(\partial_{vv}\Phi(u,v)\cdot\right.\\
&\left. \cdot\frac{(u-v)(v+\nu)}{2(u+\nu)} -
\frac{8v^2-8uv+8v\nu-2u\nu+3\nu^2+3u^2}{8(u+\nu)(v+\nu)(u-v)}(3t+\Phi(u,v))
\right)
+O(\delta^3),\\
\displaystyle 0 =&
\frac{(2v-u+\nu)(3t+\Phi(u,v))-2(u-v)(v+\nu)\partial_v\Phi(u,v)}{2(u+\nu)}
+O(\delta^2).
\end{split}
\end{equation}
From the above, we deduce that, in the limit $\delta\to 0$, the
hodograph transform (\ref{abe:hodosol}) reduces to the form
\begin{equation}
\label{delta0} \left\{
\begin{split}
0 &= (3u(t) +2\nu) t +f_-(u)-x,\\
0 &= (u-v) (t+\partial_uQ(u,v)) +(v+\nu) (3t+\Phi(u,v)),\\
0 &= (u-2v-\nu) (\Phi(u,v) +3t) +2(v+\nu)(u-v)\partial_v \Phi(u,v).
\end{split}
\right.
\end{equation}
The above system of equations enables one to determine $x$, $u$
and $v$ as functions of time. We denote this time dependence as
$x=x^-(t)$,  $v=v(t)$ and $u=u(t)$.
Observe that $u(t)=u(x^-(t),t)$ where $u(x,t)$ is the solution of the Hopf equation. In what follows we will always denote by $u$ or $u(t)$ the solution of the Hopf equation $u(x,t)$ at the leading edge $x=x^-(t)$ while we  will refer to the solution of the CH equation as $u(x,t,\epsilon)$.
 The derivatives with respect to
time of these quantities are given by
\begin{equation}\label{ut}
\begin{split}
x^-_t &= 3u+2\nu+(3t+f^{\prime}(u))u_t, \\
u_t &=
-2\frac{(u-v)(u+2v+3\nu)}{(u+\nu)(3t+f^{\prime}_-(u))},\\
v_t &= -\frac{4(u-3\nu-4v)(v+\nu)(u-v)}{V},
\end{split}
\end{equation}
with
\begin{equation}
\label{V}
V =
4(v+\nu)^2(u-v)^2\partial_{vv}\Phi(u,v)-(3t+\Phi(u,v))(8v\nu+3u^2-8uv-2u\nu+3\nu^2+8v^2).
\end{equation}
We are interested in studying the behaviour of the one--phase
solution (\ref{abe:ellisol}) near the leading edge, namely when
$0<x-x^-(t)\ll 1$. To this aim, we introduce two unknown functions
of $x$ and $t$,
\[
\delta= \delta (x-x^-(t)),\quad\quad\quad \Delta=
\Delta(x-x^-(t)),
\]
which tend to zero as $x\to x^-(t)$. We now derive the dependence
of $\Delta$ as a function of $x-x^-(t)$. Let us fix
\begin{equation}
\label{dD} \beta_2 = v+\delta, \quad\quad \beta_3 = v-\delta,\quad\quad
\beta_1 =u+\Delta,\quad\quad \Delta\to 0.
\end{equation}
For the first equation in (\ref{abe:hodonodeg}) near $\delta=0$,
we find
\[
x = (3\beta_1+2\nu)t + f_-(\beta_1)-\frac{\delta^2}{2}
\frac{(3t+\Phi(\beta_1,v))(2v+\beta_1+3\nu)}{(\beta_1-v)^2}
+O(\delta^3). \] We then substitute $\beta_1 =u+\Delta$ in the
above equation, we insert (\ref{abe:quv}), (\ref{ident}) and
(\ref{delta0}) into it, and we get
\begin{equation}
\label{Deltadep} x-x^-(t)\approx \Delta (f^{\prime}_-(u) +3t)
-\frac{\delta^2}{4} \frac{(u+2v+3\nu)(x-x^-(t))}{(u-v)^2(v+\nu)},
\end{equation}
so that
\begin{equation}
\label{Deltax} \Delta \approx \frac{x-x^-(t)}{3t+f^{\prime}_-(u)}.
\end{equation}
Similarly, using the second equation in (\ref{abe:hodonodeg}), we
arrive at
\begin{equation}\label{deltax}
x-x^-(t) \approx \delta^2 c,\quad\quad c=
-\frac{V}{8(u+\nu)(v+\nu)(u-v)}>0,
\end{equation}
with $V$ defined in (\ref{V}).

\begin{theorem}
In the limit
\[
\beta_2 = v+\delta, \quad\quad \beta_3 = v-\delta,\quad\quad \beta_1=u+\Delta,
\quad\quad \delta,\Delta\to 0,
\]
the one-phase solution of the CH equation, implicitly defined by
(\ref{abe:ellip}),  has the following trigonometric expansion
\begin{equation}\label{ellilim}
u(x,t,\epsilon) \approx u(t) +\frac{x-x^-(t)}{3t+f^{\prime}_-(u)}
+2\delta \cos\left(\frac\xi{\epsilon}\right)
-\frac{\delta^2}{2} \frac{u+\nu}{(u-v)(v+\nu)} \left( 1
-\cos\left(2\frac\xi{\epsilon}\right)\right),
\end{equation}
with
\begin{equation}\label{zeta}
\begin{split}
&\xi=\xi_0 + \xi_1+O((x-x_-(t))^2),\\
&\xi_0 =-2\sqrt{(u-v)(v+\nu)} (\Phi(u,v) +3t) \quad\quad
\xi_1 =\sqrt{\frac{u-v}{v+\nu}}(x-x_-(t)).
\end{split}
\end{equation}

\end{theorem}

{\sl Proof:} We first prove (\ref{zeta}). Let $\zeta
=\xi/\epsilon$ in (\ref{abe:ellip}), so that
\[
\xi = k \left[ x-(\beta_1+\beta_2+\beta_3+2\nu)t
-q(\beta_1,\beta_2,\beta_3)\right],
\]
with $k$ as is (\ref{abe:kappa}), $q$ as in (\ref{q}).
In the limit $\delta,\Delta\to 0$, with $\beta_2 = v+\delta$, $\beta_3 =
v-\delta$, $\beta_1 = u+\Delta$, we get
\[
\begin{split}
\xi &= \sqrt{\frac{u-v}{v+\nu}}  \left( 1
+\frac{\Delta}{2(u-v)}\right) \left( 1-\frac{\delta^2}{16}
\frac{(u+\nu)(3\nu+4v-u)}{(v+\nu)^2(u-v)^2}\right)
\bigg(x-\\
&\left. -(u+2\nu+2v+\Delta)t-Q(u,v)-\Delta \partial_u Q(u,v)
-\frac{\delta^2}{4}\partial_{vv} Q(u,v)\right) + O(\delta^4),
\end{split}
\]
where we use $\Delta = O(\delta^2)$ to estimate the error in the
formula above. Then, we insert the solution $x_-(t)$ of equations
(\ref{delta0}), we use (\ref{Deltax}) and (\ref{deltax}) to
estimate $\Delta$ and $\delta^2$ and we get
\[
\xi = -2\sqrt{(u-v)(v+\nu)} (\Phi(u,v) +3t) +
\sqrt{\frac{u-v}{v+\nu}}(x-x_-(t)) + O((x-x_-(t))^2),
\]
where $\Phi$ is defined in (\ref{abe:quv}), which coincides with (\ref{zeta}).

We now prove that in the limit $\delta\to 0$, $u(x,t,\epsilon)$,
defined implicitly as a function of $\zeta$ in (\ref{abe:ellip}),
has the following trigonometric series expansion
\begin{equation}\label{abe:udelta}
u(x,t,\epsilon) = \beta_1 +2\delta \cos(\zeta) -\frac{\delta^2}{2}
\frac{\beta_1+\nu}{(\beta_1-v)(v+\nu)} \left( 1 -\cos(2\zeta)\right) +
O(\delta^3).
\end{equation}
To this aim, we expand $u(x,t,\epsilon)$ and $\zeta$ defined in
(\ref{abe:ellip}) in trigonometric series of $z$ and we compute
$z$ as a  function of $\zeta$.

We need the following series expansions for the Jacobi theta
functions  \cite{WW}
\begin{equation}\label{abe:thetaexp}
\begin{split}
\displaystyle \log \left( \frac{\vartheta_3 (z-p)}{\vartheta_3
(z+p)}\right) &= 4\sum_{n=1}^{+\infty}
\frac{(-1)^{n+1}}{n}\frac{\q^n}{1-\q^{2n}}\sin(2n\pi z)\sin(2n\pi p),\\
\displaystyle \frac{\vartheta^{\prime}_3 (z\pm p)}{\vartheta_3
(z\pm p)} &= 4\pi\sum_{n=1}^{+\infty} (-1)^n
\frac{\q^n}{1-\q^{2n}}\sin(2n\pi (z\pm p)),
\end{split}
\end{equation}
where $p$ is as in (\ref{abe:p0}) and
\begin{equation}\label{epit}
\q= \exp\left( -\pi K^{\prime}(s)/K(s)\right) =
\frac{(\beta_1+\nu)}{8(v+\nu)(\beta_1-v)}\delta +O(\delta^3).
\end{equation}
Inserting the above expansion and
\[
\begin{split}
\sin(2\pi p) &= 2\frac{\sqrt{(v+\nu)(\beta_1-v)}}{\beta_1+\nu}+O(\delta^2),\\
\sin(4\pi p ) &= -4
\frac{(2v+\nu-\beta_1)\sqrt{(v+\nu)(\beta_1-v)}}{(\beta_1+\nu)^2}+
O(\delta^2),
\end{split}
\]
into (\ref{abe:thetaexp}), we get
\[
\begin{split}
u(x,t,\epsilon) &= \beta_1 +2\delta \cos(2\pi z) +\frac{\delta^2}{2}
\frac{\beta_1-\nu-2v}{(v+\nu)(\beta_1-v)} \left( 1 -\cos(4\pi z)\right)+
O(\delta^3),\\
\zeta &= 2\pi z -\frac{\delta}{v+\nu} \sin(2\pi z)+O(\delta^2)\end{split}
\]
from which we find
\[
2\pi z = \zeta +\frac{\sin(\zeta)}{v+\nu}\delta +O(\delta^2),
\]
and the assertion easily follows. $\square\quad$

\medskip

\section{Painlev\'e equations at the leading edge\label{2}}
In this section we propose a multiscale description of the
oscillatory behavior of the solution to the CH equation in the small
dispersion limit ($\epsilon\to 0$) close to the leading edge
$x_-(t)$ where $\beta_2=\beta_3=v$ and $\beta_1=u$. We follow
closely the approach \cite{grakle2} for the corresponding KdV
situation. The ansatz for the multiscale expansion to the CH
solution close to the leading edge is inspired by the asymptotic
solution in the Whitham zone discussed in the previous section.
Numerically we find that the quantity $\delta$ in (\ref{ellilim})
is of the order $\epsilon^{1/3}$. This implies with (\ref{deltax})
that $x-x_-(t) \approx \epsilon^{2/3}$ in the double scaling limit
$\epsilon\to 0$ and $x\to x_-(t)$. We are thus led to introduce
the rescaled coordinate $y$ near the leading edge,
\begin{equation}
\label{y}
y = \epsilon^{-2/3} (x-x_-(t)),
\end{equation}
which transforms the CH equation (\ref{CH}) to the form
\begin{equation}
\label{CHy} (3u+2\nu-x_t^-)u_y +\epsilon^{2/3} (u_t
-2u_yu_{yy}-(u-x^-_t)u_{yyy}) -\epsilon^{4/3} u_{yyt} =0,
\end{equation}
where $x_t^-=\frac{d}{dt}x_-(t).$ 

Numerically the corrections to the Hopf solution near the leading
edge are of order $\epsilon^{1/3}$ and thus we make as in
\cite{grakle2} the ansatz
\begin{equation}\label{CHexp}
u(y,t,\epsilon) = U_0(y,t) + \epsilon^{1/3} U_1(y,t) +
\epsilon^{2/3} U_2(y,t) + \epsilon U_3 (y,t)+ \dots,
\end{equation}
where $U_0=u(t)$ is the solution of the Hopf equation at the leading edge. We assume
that the terms $U_k$, $k\ge 1$ contain oscillatory terms of the
order $1/\epsilon$ and take the form
\begin{equation}\label{Ui}
\begin{split}
U_1 (y,t) &= a(y,t) \cos \left(
\frac{\psi(y,t)}{\epsilon}\right),\\
U_2 (y,t) &= b_1 (y,t) +  b_2(y,t)\cos \left(
\frac{2\psi(y,t)}{\epsilon}\right), \\
U_3 (y,t) &= c_0 (y,t) + c_1(y,t) \cos \left(
\frac{\psi(y,t)}{\epsilon}\right) + c_2(y,t) \sin \left(
\frac{2\psi(y,t)}{\epsilon}\right) + \\
& + c_3 (y,t) \cos \left( \frac{3\psi(y,t)}{\epsilon}\right),
\end{split}
\end{equation}
where
\begin{equation}\label{psi}
\psi (y,t) = \psi_0(y,t) + \epsilon^{1/3} \psi_1 (y,t) +
\epsilon^{2/3} \psi_2 (y,t) + \epsilon \psi_3 (y,t) +\dots .
\end{equation}
Terms proportional to $\sin (\psi/\epsilon)$ in $U_{1}$ can be absorbed
by a redefinition of $\psi$, the higher order terms are chosen to
compensate terms of lower order appearing in the solution to CH due
to the non-linearities. We immediately find $\partial_y
\psi_0 =0=\partial_y \psi_1$ from the terms of order
$\epsilon^{-2}$ and $\epsilon^{-1}$. The term of order
$\epsilon^0$ gives
\begin{equation}\label{eq13}
\left( 1 + (\partial_y \psi_2)^2 \right) \partial_t \psi_0 +
\partial_y \psi_2 \left( 3u +2\nu -x^-_t \right) + \left( u(t)-x^-_t \right)
(\partial_y \psi_2)^3=0.
\end{equation}
In particular, (\ref{eq13}) is algebraic in $\partial_y \psi_2$
with coefficients depending on $t$ only, i.e.,
$\psi_2(y,t)=f_2(t)y +f_1(t)$, with $f_{1,2}(t)$ to be determined.  At the order $\epsilon^{1/3}$, we
get $\partial_t \psi_1=0$ and
\begin{equation}\label{eq14}
a^2 \left( 1 +f_2^2 \right)^2 = 8b_2 f_2^2 \left( u+\nu\right),
\end{equation}
\begin{equation}\label{eq15}
f_2^4 (x^-_t -u) +2f_2^2 (x^-_t +\nu) +x^-_t -2\nu -3u=0.
\end{equation}
At the order $\epsilon^{2/3}$, using (\ref{eq13}) to (\ref{eq15}), we
find $\partial_y \psi_3 = 0$ and
\begin{equation}\label{c2-3}
c_2 = \frac{a\partial_y a \left(f_2^4-1\right)
}{4(u+\nu)f_2^3},\quad\quad c_3 = \frac{a^3 \left( 3 +
7f_2^2\right)\left( 1+ f_2^2\right)^3 }{256 (u +\nu)^2f_2^4},
\end{equation}
\begin{equation}\label{b1}
b_1 = -\frac{1}{8} \frac{(a^2 (3+f_2^2)+4 u_t y)
(1+f_2^2)^2}{f_2^2 (3+f_2^2) (u+\nu)}+g_1(t),
\end{equation}
with $g_1(t)$ integration constant,
\begin{equation}\label{P2-mod}
\begin{split}
\partial_{yy} a &-\frac{\left( 1 + f_2^2\right)^4 }{32(u+\nu)^2 f_2^2}a^3 +
\frac{ay}{4}\frac{u_t\left(1+f_2^2\right)^4}{f_2^2
(-3+f_2^2) (u+\nu)^2)}\\
&-\frac{a}{2}\frac{ (1+f_2^2)^3
\partial_t\psi_2}{f_2 (u+\nu) (-3+f_2^2)}
-\frac{g_1a}{2} \frac{(3+f_2^2) (1+f_2^2)^2}{(u+\nu)
(-3+f_2^2)}=0.
\end{split}
\end{equation}

We determine $\partial_y \psi_2$ inserting (\ref{ut}) inside
(\ref{eq15}), so that
\begin{equation}\label{psi2pm}
\partial_y \psi_2^2 = f_2^2 \in \left\{ \frac{u-v}{v+\nu},
\frac{2v+u+3\nu}{u-\nu-2v}\right\}, \end{equation} where the sign
in front of the square root has to be chosen in such a way that
the r.h.s. is positive. Then comparing (\ref{psi2pm}), (\ref{psi}),
(\ref{zeta}) and (\ref{y}), we conclude that
\begin{equation}\label{psi2}
\psi_2 (y,t) = y \sqrt{\frac{u-v}{v+\nu}}.
\end{equation}
Inserting (\ref{psi2}) into (\ref{eq13}), we find
\begin{equation}\label{psi0t}
\partial_t \psi_0(t) =
-4\sqrt{\frac{v+\nu}{u-v}}\frac{(u-v)^2}{u+\nu}.
\end{equation}
By comparison with (\ref{zeta}), we observe that
\[
-2\frac{d}{dt} \left( \sqrt{(u-v)(v+\nu)} (\Phi(u,v) +3t) \right) =
-4\sqrt{\frac{v+\nu}{u-v}}\frac{(u-v)^2}{(u+\nu)}.
\]
Then integrating the l.h.s.\  of the above expression between $t_c$ and $t$ we arrive at  (\ref{zeta}), namely
\begin{equation}\label{psi0}
\psi_0(t) = -2\sqrt{(u-v)(v+\nu)} (\Phi(u,v) +3t).
\end{equation}
Moreover, consistency between (\ref{CHexp}) and (\ref{ellilim})
implies that
\begin{equation}\label{d-eps}
\delta = \frac{\epsilon^{1/3}a}{2},
\end{equation}
from which we immediately verify that
\[
\epsilon^{2/3} b_2 = \frac{\delta^2}{2}
\frac{u+\nu}{(u-v)(v+\nu)},
\]
\[
\epsilon^{2/3} b_1 = \frac{x-x_-(t)}{3t+f^{\prime}(u)}
-\frac{\delta^2}{2} \frac{u+\nu}{(u-v)(v+\nu)},
\]
with $g_1(t)\equiv 0$ in (\ref{b1}).

Summarizing, we get
\begin{equation}\label{umulti}
u(x,t,\epsilon) = u(t) + \epsilon^{1/3} a
\cos\left(\frac{\psi}{\epsilon}\right) + \epsilon^{2/3}
\left[\frac{a^2(u+\nu)( \cos(2\psi/\epsilon)-1)}{8(u-v)(v+\nu)}
+\frac{y}{3t+f^{\prime}(u)}\right]+O(\epsilon),
\end{equation}
where
\begin{equation}
\psi(y,t)
=-2\sqrt{(u-v)(v+\nu)} (\Phi(u,v) +3t)+ \epsilon^{2/3}y \sqrt{\frac{u-v}{v+\nu}}+O(\epsilon),
\end{equation}
with $Phi(u,v)$ defined in (\ref{abe:quv})
and $a$ satisfies the Painlev\'e-II equation
\begin{equation}\label{P2-oldvar}
\partial_{yy} a - A_1 ay -A_2a^3=0,
\end{equation}
with
\begin{equation}
A_1=\frac{(u+\nu)^3v_t}{4(u-v)(v+\nu)^3(3\nu+4v-u)} \quad\quad A_2
= \frac{(u+\nu)^2 }{32(u-v)(v+\nu)^3 }.
\end{equation}
Using (\ref{ut}), we get
\[
A_1 =\frac{(u+\nu)^3}{V(v+\nu)^2}.
\]
with $V$ defined in (\ref{V}).
Making the substitution $z=\alpha y$, $a=\beta A$ with
\[
\alpha = \left( A_1\right)^{1/3},\quad\quad \beta
=\frac{\alpha\sqrt{2}}{\sqrt{A_2}}, \]
 we arrive to the special Painlev\'e II equation in normal form
\begin{equation}\label{P2}
A_{zz} =zA+2A^3.
\end{equation}
For $x-x_-(t)>0$, from the small amplitude limit of the one--phase
solution to the CH equation we get
\[
a = 2\epsilon^{-1/3} \delta \approx
2\epsilon^{-1/3}\sqrt{\frac{x-x_-}{c}},
\]
with $c$ defined in (\ref{deltax}),
so that in the limit $\epsilon\to 0+$ (equivalently $y\to +\infty$
or $z\to -\infty$), we have
\[
A(z) \approx \sqrt{-z/2}.
\]
If $x\ll x_-(t)$, the CH solution is approximately the solution to
the Hopf equation so that $a(y)\approx 0$, for $y\to-\infty$
(equivalently $z\to+\infty$). Therefore we conclude that
\[
\lim_{z\to+\infty} A(z)=0.
\]
The solution to the Painlev\'e II equation satisfying such
asymptotic condition exists and is unique and pole free on the
real line according to Hastings and McLeod \cite{HMcL}.

\section{Numerical solution of CH and Whitham equations}%
\label{3}
In this section we will solve numerically  the CH and the Whitham
equations for initial data in the Schwartzian class of rapidly
decreasing functions with a single negative hump. As a concrete example we
will study the initial datum
\begin{equation}
    u_{0}:=u(x,0) = -\mbox{sech}^{2}x
    \label{initial}.
\end{equation}
For this initial datum the non--breaking condition (\ref{nonbreak}) is satisfied if $\epsilon^2<(\nu-1)/2$.

\subsection{Numerical solution of the CH equation}
The resolution of the rapid modulated oscillations in the
region of a dispersive shock is numerically demanding. The strong
gradients in the oscillatory regions require efficient approximation
schemes which do not introduce an artificial numerical dissipation
into the system.  We therefore use Fourier spectral methods which
are known for their excellent approximation properties for smooth
functions whilst minimizing the introduction of numerical viscosity.
We restrict to initial data, where $u_{0}-\epsilon^{2}u_{0,xx}+\nu$
does not change sign, to ensure analyticity of the CH solutions. The 
Schwartzian solutions can be treated as effectively periodic if the
computational domain is taken large enough that the solution is of
the order of machine precision ($\approx10^{-16}$ in Matlab) at the
boundaries. We always
choose the computational domain in this way to avoid Gibbs phenomena
at the boundaries.

Even with spectral methods, a large number of Fourier modes is
needed to  resolve the rapid oscillations numerically. To obtain
also a high resolution in time, we use high order finite
difference methods, here a fourth order Runge-Kutta scheme. For
stability reasons a sufficiently small time step has to be chosen.
Unconditionally stable implicit schemes could be used instead, but
these can be only of second order. Such approaches would be too
inefficient for the  precision requested. Notice that the terms with
the highest derivative in CH (\ref{CH}) are not linear in contrast
to the KdV equation (we choose the KdV equation in a way that it
has the same dispersionless equation as (\ref{CH}), the term
proportional to $\nu$ can be always eliminated here in contrast to
CH by a Galilean transformation)
\begin{equation}
    u_{t}+(3u+2\nu)u_{x}+\epsilon^{2}u_{xxx}=0.
    \label{KdV}
\end{equation}
For such equations efficient integration schemes are known. In
\cite{etna} it was shown that \textit{exponential time
differentiation} (ETD) methods \cite{coxma,minwri} are the most efficient in the
small dispersion limit of KdV. For CH such an approach is not possible
due to the nonlinearity of the highest order derivatives. But we can
use analytic knowledge of the solution: The
dispersionless equation, the Hopf equation, will have a point of
gradient catastrophe at the critical time $t_{c}$, for the example
(\ref{initial}) $t_{c}=\sqrt{3}/4\approx 0.433$. For times $t\ll
t_{c}$, the CH solution is very close to the Hopf solution with
moderate gradients. In this case we can use a larger time step.
For  times close to the critical time and beyond (in our
example $t>0.4$), we have to use considerably smaller time steps.

\begin{figure}[!htb]
\centering
\epsfig{figure=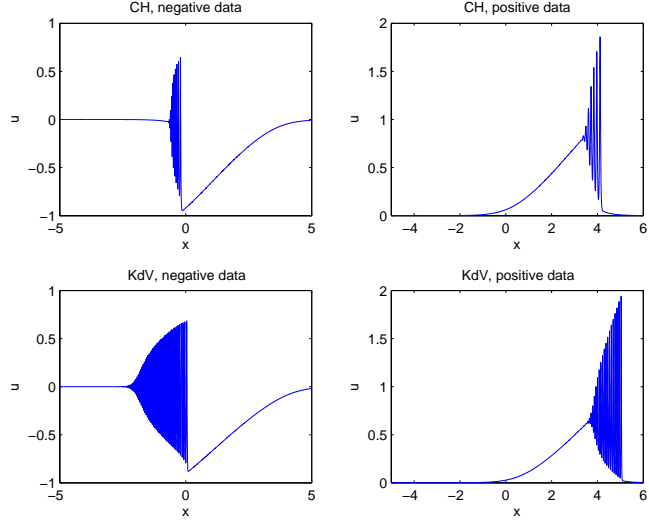, width=10cm}
\caption{Solution to the CH equation (\ref{CH}) and KdV equation
(\ref{KdV})  with $\nu=1.2$ for the initial
data $u_{0}=-\mbox{sech}^{2}x$ on the left side
and $u_{0}=\mbox{sech}^{2}x$ on the right side at $t=1$ for
$\epsilon=10^{-2}$.}
\label{KdVCH}
\end{figure}

The main difference to the KdV equation is the
non-locality of CH, which provides some filtering for the high
frequencies: if we denote the Fourier transform of $u$ with respect
to $x$ by
$\hat{u}(k,t):=\int_{-\infty}^{\infty}u(x,t)e^{ikx}dx$,
we get for (\ref{CH}) in Fourier space
$$
\hat{u}_{t}=\frac{1}{1+\epsilon^{2} k^{2}}\left(
\frac{3}{2}ik\widehat{u^{2}}+2\nu ik\hat{u}-
    \epsilon^{2}(2\widehat{u_{x}u_{xx}}+\widehat{uu_{xxx}})\right).
$$
The term $1/(1+\epsilon^{2} k^{2})$ in the above equation is the reason
why CH gives for large spatial frequencies a better approximation to
one-dimensional wave phenomena
than KdV with a term proportional to $k^{3}$ in Fourier space.
Its effect in the present context is twofold: First it
provides a high frequency filtering which allows in practice for
larger time steps in the computation. Secondly it suppresses the
rapid modulated oscillations in the shock region of the
dispersionless equation. This can be seen in Fig.~\ref{KdVCH}, where
solutions to KdV and CH for the same initial data are shown.
The KdV solutions always show much more
oscillations than the CH solutions.

\begin{figure}[!htb]
\centering
\epsfig{figure=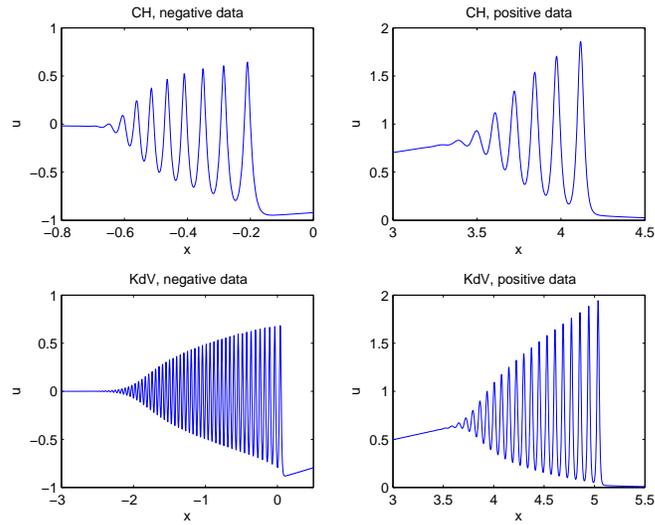, width=10cm}
\caption{Magnification of the oscillations in Fig.~\ref{KdVCH}.}
\label{KdVCH2}
\end{figure}

Notice that despite the lower number of oscillations in the CH
solutions in Fig.~\ref{KdVCH}, depending on the value of $\nu$ a
higher number of Fourier modes is needed to resolve the
oscillations. This is due to the fact that the CH solution is less smooth
than the KdV solution and thus less localized in Fourier space,
as can be seen in Fig.~\ref{KdVCH2}.
Therefore, to treat the same values of the small dispersion
parameter $\epsilon$ for CH as for KdV, we need higher  temporal
and spatial resolution.
It is thus computationally more demanding to obtain the same
number of oscillations in CH solutions as in   KdV
solutions, the latter being important to obtain a valid statistics
for the scaling studied below.

The quality of the numerics is
controlled via energy conservation for CH,
$$
E \sim 2\nu u^{2}+\epsilon^{2}u u_{x}^{2}+u^{3}.$$
The numerically computed energy will depend on time due to numerical
errors. As was discussed in \cite{etna}, energy conservation can
thus be used to check numerical accuracy. In practice energy conservation
overestimates numerical precision by 1-2 order of magnitude. We
typically solve the CH equation with a relative numerical error
$\Delta E/E=10^{-6}$. This ensures that the difference between the
numerical and the asymptotic CH solution, which is typically of the
order of $\epsilon$ or larger, is entirely due to the asymptotic
description.

\subsection{Numerical solution to the Whitham equations}
\begin{figure}[!htb]
\centering \epsfig{figure=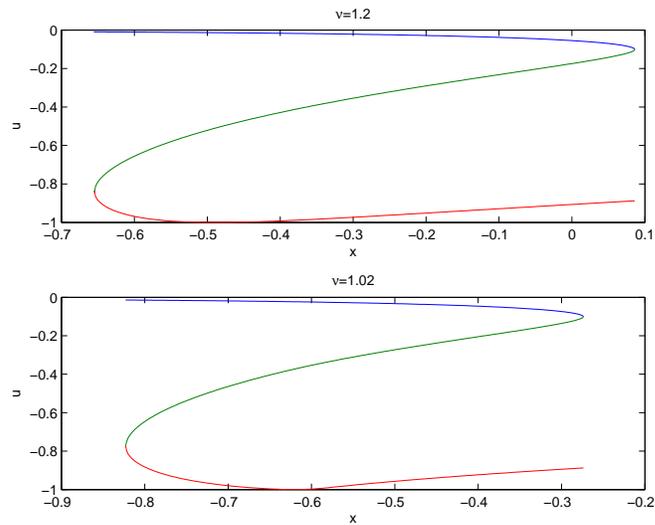, width=10cm}
\caption{Solutions $\beta_{i}$, $i=1,2,3$ to the Whitham equations
(\ref{abe:whi}) for $u_{0(x)}=-\mbox{sech}^{2}x$ and $t=1$ for two
values of $\nu$.} \label{Whithb}
\end{figure}

The Whitham equations (\ref{abe:whi}) have a similar form as the
respective equations for KdV. We use the same procedure to solve them
numerically: we first solve the equations at the edges of the Whitham
zone and then at intermediate points. For details the reader is
referred to \cite{grakle2}. Typical solutions for the Whitham
equations can be seen in Fig.~\ref{Whithb}. In the shown example, the
quantity $\beta_{3}$  crosses the hump at $-1$.

\begin{figure}[!htb]
\centering
\epsfig{figure=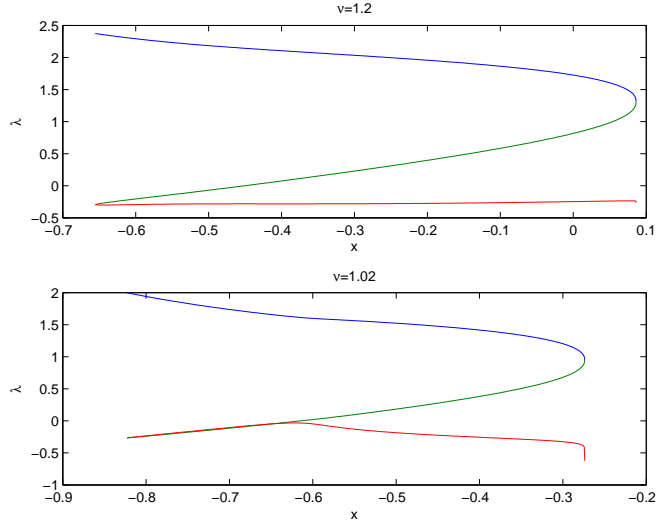, width=10cm}
\caption{The speeds $C_{i}$, $i=1,2,3$ (\ref{Ci1}) for
$u_{0}(x)=-\mbox{sech}^{2}x$ and $t=1$ for two values of $\nu$.}
\label{chspeeds}
\end{figure}

\begin{figure}[!htb]
\centering
\epsfig{figure=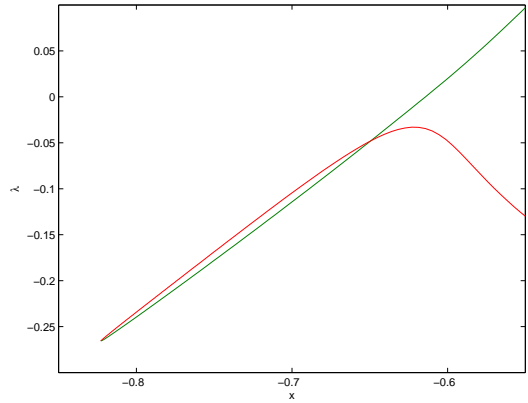, width=8cm}
\caption{Magnification of the lines in the lower figure of
Fig.~\ref{chspeeds} close to the crossing of the lines.}
\label{chspeedsm}
\end{figure}

In contrast to KdV, the system (\ref{abe:whi}) is not strictly
hyperbolic, i.e., the speeds $C_{i}$, $i=1,2,3$, in (\ref{Ci1})
do not satisfy for all $x$ and $t>t_{c}$ the relation
$C_{1}>C_{2}>C_{3}$. In fact the lines
$C_{i}(x)$ can cross for a given time for $\nu\sim1$ as can be
seen in Fig.~\ref{chspeeds}.
For $\nu=1.02$ the speeds $C_{2}$ and $C_{3}$ intersect
in the shown example in the interior of the Whitham zone as can be
seen in more detail in Fig.~\ref{chspeedsm}.
This behavior of the speeds has no influence on the numerical
solubility of the Whitham equations. It also does not influence the
quality of the asymptotic solution in these cases as can be seen
in Fig.~\ref{chasym2nu}.

\subsection{Quantitative comparison of the CH solution and the
asymptotic solution} The asymptotic description of the small
dispersion limit of the CH equation is as follows: for times
$t<t_{c}$, the Hopf solution for the same initial datum provides
an asymptotic description. For $t>t_{c}$, the Whitham zone opens.
Outside this zone, the Hopf solution
again serves as an asymptotic solution.
In the interior the one-phase solution to the CH equation  describes the oscillatory
behavior. It is given on an
elliptic curve with branch points being solutions of
the Whitham equations.

Below we will study the
validity of this asymptotic description in various regions of the
($x,t$)-plane. To study the $\epsilon$ dependence of a certain
quantity  $A$, we perform a linear regression analysis for the
dependence of the logarithms, $\ln A=a\ln \epsilon +b$. We compute
all studied quantities for the
$\epsilon$ values $\epsilon=10^{-\alpha}$ with $\alpha\in
[1,1.25,1.5,\ldots,3]$. Generally it is found that the correlations and
the standard deviations are worse than in the KdV case
due to the lower number of oscillations.

\textit{Before breakup,} $t\ll t_{c}:$\\
For times much smaller than the critical time, we find that the
$L_{\infty}$ norm of the difference between Hopf and CH
solution decreases as $\epsilon^{2}$. More precisely we find by
linear regression an exponent $a=1.91$ with correlation coefficient
$r=0.999$ and standard deviation $\sigma_{a}=0.06$.

\textit{At breakup,} $t\sim t_{c}:$\\
For times close to the breakup time, the Hopf solution develops a
gradient catastrophe. The largest difference between Hopf and CH
solution can be found close to the breakup point. We determine the
scaling of the $L_{\infty}$ norm of the difference between Hopf
and CH solution on the whole interval of computation. We find that
its scaling is compatible with $\epsilon^{2/7}$ as conjectured in
\cite{Du}. More precisely we find in a linear regression
analysis $a=0.28$ ($2/7=0.2857\ldots$) with a correlation
coefficient $r=0.998$ and standard deviation $\sigma_{a}=0.015$. An
enhanced asymptotic description of the CH solution near the
breakup point in terms of a solution to the Painlev\'e I2 equation
was conjectured in \cite{Du} and studied numerically in
\cite{grakle12}.

\begin{figure}[!htb]
\centering
\epsfig{figure=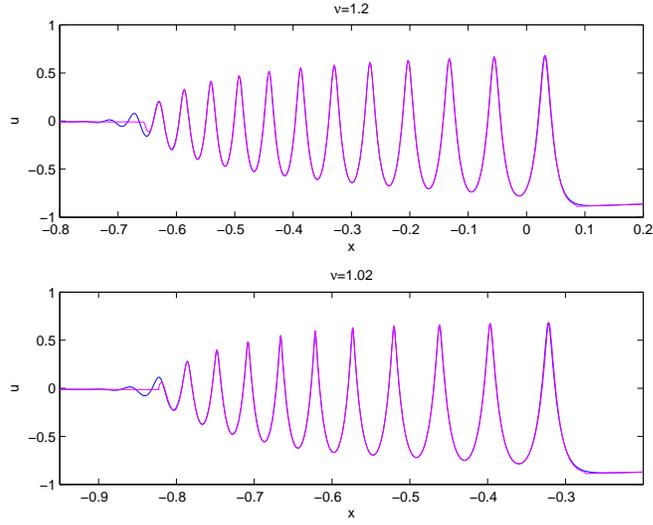, width=10cm}
\caption{CH solution (blue) and asymptotic solution (magenta)
for the initial datum $u_{0}=-\mbox{sech}^{2}x$ and $\epsilon=10^{-2}$
for $t=1$ and two values of $\nu$}
\label{chasym2nu}
\end{figure}

\begin{figure}[!htb]
\centering \epsfig{figure=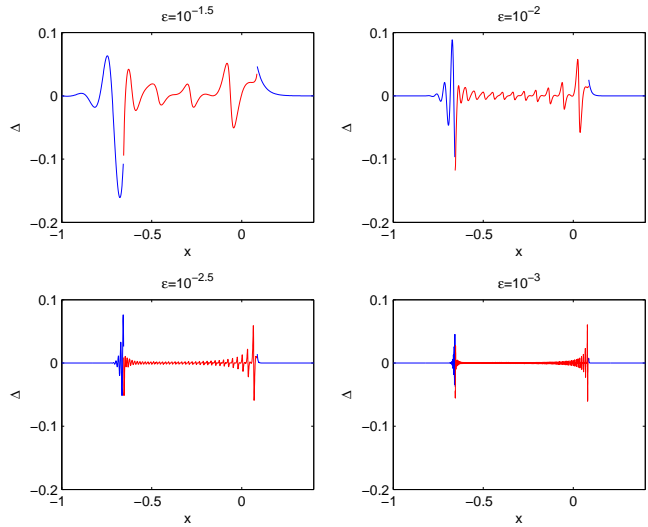, width=10cm}
\caption{Difference of the CH solution and the asymptotic solution
for the initial datum $u_{0}=-\mbox{sech}^{2}x$ and $\nu=1.2$ for
$t=1$ for several values of $\epsilon$. In the Whitham zone (red),
the asymptotic solution is given in terms of the one-phase CH
solution solution, outside by the solution to the Hopf equation
with the same initial datum.} \label{chdelta4e}
\end{figure}

\textit{Times } $t>t_{c}$:\\
For times $t\gg t_{c}$ it can be seen from Fig.~\ref{ch1e3asym} and
Fig.~\ref{chasym2nu} that the asymptotic solution gives a very
satisfactory description of the oscillations except at the boundaries
of the Whitham zone.
The asymptotic solution is so close to the approached solution that
one can only see discrepancies near the boundary of the Whitham zone,
where the asymptotic solution is just $C^{0}$. Thus one
has to consider the difference between the solutions as shown in
Fig.~\ref{chdelta4e}. The quality of the numerics allows the study of
the scaling behavior at various points in the Whitham zone.

\textit{Interior of the Whitham zone:}\\
If we study the $\epsilon$-dependence of the $L_{\infty}$-norm of
the difference near the middle of the Whitham zone (we take the
maximum of this difference near the geometric midpoint), we find
that the norm scales as $\epsilon$. More precisely we find an
exponent $a=0.9$ with correlation coefficient $r =0.999$ and
standard deviation $\sigma_{a}=0.04$. There is obviously a certain
arbitrariness in our definition of this error especially for
larger values of $\epsilon$, where there are only few
oscillations. In these cases the errors are read off close to the
boundaries of the Whitham zone where much smaller exponents for
the error are observed (see below). This scaling gives nonetheless
strong support for the conjectured form of the phase of the
one-phase solution in the Whitham zone, since even a small
analytical error in the phase would lead to large errors which
would not decrease with $\epsilon$.

\textit{Leading edge of the Whitham zone:}\\
Oscillations can always be found
outside the Whitham zone, whereas the Hopf solution does not show any
oscillations. The biggest difference always occurs at the boundary of the Whitham
zone. It scales as $\epsilon^{1/3}$. More precisely we find in the
Hopf zone $a=0.33$ with $r=0.99$ and $\sigma_{a}=0.04$. In the
interior of the Whitham zone, $a$ has a similar value, but the
correlation is worse. If one studies the scaling of the zone, where
the difference between Hopf and CH solution near the leading edge has
absolute value larger than some treshold (we use $10^{-4}$), we find a
decrease compatible with $\epsilon^{2/3}$, more precisely  $a=0.81$
with $r=0.99$ and $\sigma_{a}=0.07$.

\textit{Trailing edge of the Whitham zone:}\\
The biggest difference is always found at the boundary of the Whitham
zone. Its scaling is compatible  with $\epsilon^{1/2}$. We find
$a=0.51$ with $r=0.9997$ and $\sigma_{a}=0.01$.

\section{Numerical study of the multiscales expansion for the CH equation}%
\label{4}

In this section we study numerically the multiscales solution to
CH derived in  section~\ref{2}.  It will be shown that the latter
provides a better description of the asymptotic behavior near the
leading edge of the Whitham zone than the Hopf or the one-phase
solution to CH as discussed in section \ref{3}. We identify the
zone, where the latter gives a better description of CH than the
former and study the $\epsilon$-dependence of the errors. For the
numerical computation of the Hastings-McLeod solution and a
comparison to the KdV case, we refer the reader to \cite{grakle2}.

\begin{figure}[!htb]
\centering \epsfig{figure=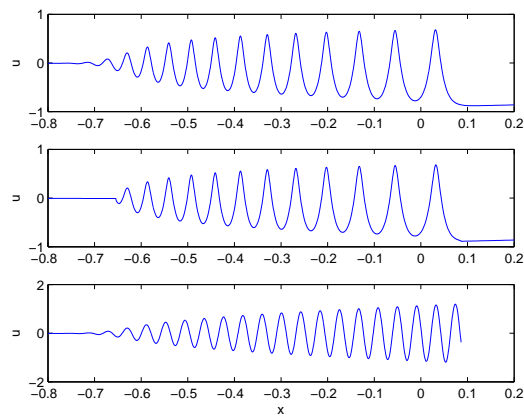, width=8cm} \caption{The
figure shows in the upper part the numerical solution to the CH
equation for the initial datum $u_{0}=-\mbox{sech}^{2}x$ and
$\nu=1.2$, $\epsilon=10^{-2}$ at $t=1$, in the middle the
corresponding asymptotic solution in terms of Hopf and one-phase
solution, and in the lower part the multiscales solution.}
\label{ch3in1}
\end{figure}

In Fig.~\ref{ch3in1} we show the CH solution, the asymptotic
solution via Whitham and Hopf and the multiscales solution near
the leading edge of the Whitham zone. It can be seen that the
one-phase solution gives a very good description in the interior
of the Whitham zone as discussed in section \ref{3}, whereas the
multiscales solution gives as expected a better description near
the leading edge.

\begin{figure}[!htb]
\centering \epsfig{figure=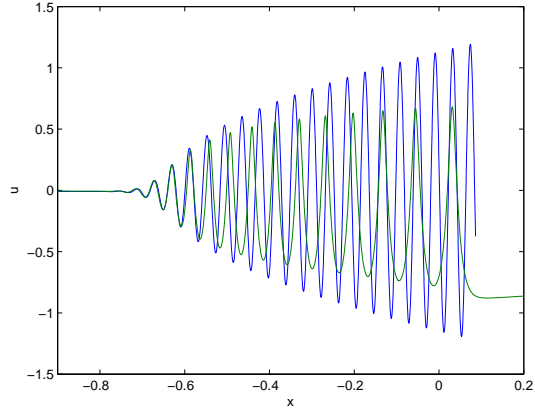, width=8cm} \caption{The
numerical solution to the CH equation for the initial datum
$u_{0}=-\mbox{sech}^{2}x$ and $\nu=1.2$, $\epsilon=10^{-2}$ at
$t=1$ in blue and the corresponding  multiscales solution in
green.} \label{chp21e4}
\end{figure}

\begin{figure}[!htb]
\centering \epsfig{figure=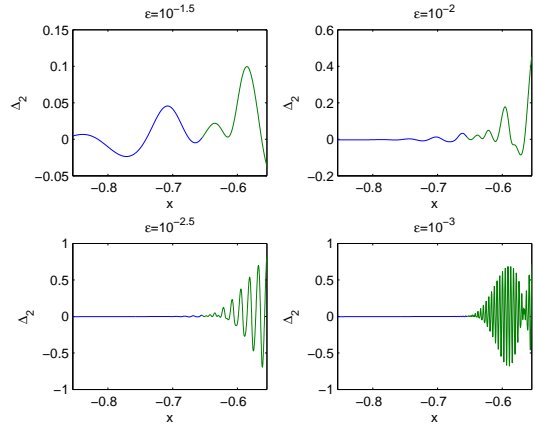, width=8cm}
\caption{The difference between the numerical solution to the CH
equation for the initial datum $u_{0}=-\mbox{sech}^{2}x$ and
$\nu=1.2$ at $t=1$ and the corresponding  multiscales solution for
4 values of $\epsilon$. The interior of the Whitham zone is shown
in green, the exterior in blue.} \label{chp21e4diff4e}
\end{figure}

In Fig.~\ref{chp21e4} the CH and the multiscales solution are
shown in one plot for $\epsilon=10^{-2}$. It can be seen that the
agreement near the edge of the Whitham zone is so good that one
has to study the difference of the solutions. The solution only
gives locally an asymptotic description and is quickly out of
phase for larger distances from the leading edge.
The difference between CH and multiscales solution is shown for
several values of $\epsilon$ in Fig.~\ref{chp21e4diff4e}. It can
be seen that the maximal error still occurs close to the Whitham
edge, but that it decreases much faster with $\epsilon$ than the
error given by the Hopf and the Whitham solution.  A linear
regression analysis for the logarithm of the difference $\Delta$
between CH and multiscales solution near the edge gives a scaling
of the form $\Delta\propto \epsilon^{a}$ with $a=0.58$ with
standard deviation $\sigma_{a}=0.17$. Since there are much less
oscillations in the CH case than in KdV, the found statistics is
considerably worse in the former case than in the latter
\cite{grakle2}, which is reflected by the low correlation coefficient
$r=0.93$ and the comparatively large standard deviation.
Nonetheless the found scaling is in accordance with the
$\epsilon^{2/3}$ scaling  expected from the multiscales expansion.

\begin{figure}[!htb]
\centering \epsfig{figure=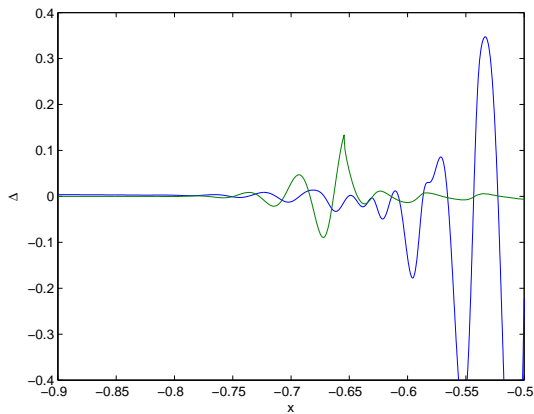, width=8cm} \caption{The
difference between the numerical solution to the CH equation for
the initial datum $u_{0}=-\mbox{sech}^{2}x$ and $\nu=1.2$,
$\epsilon=10^{-2}$ at $t=1$ for $\epsilon=10^{-2}$ and the
corresponding  multiscales solution in blue, and the difference
between CH and Hopf and one-phase solution in green.}
\label{ch1e42d}
\end{figure}

\begin{figure}[!htb]
\centering \epsfig{figure=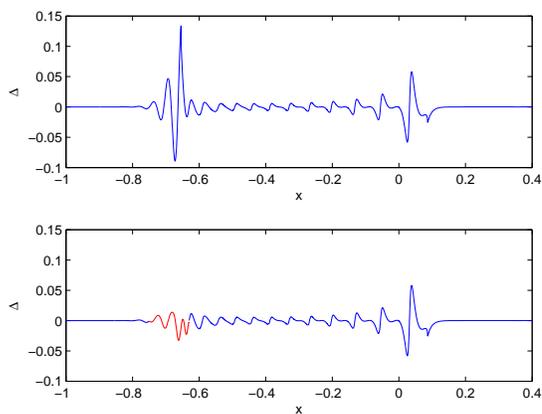, width=8cm} \caption{In
the upper part one can see the difference between the numerical
solution to the CH equation for the initial datum
$u_{0}=-\mbox{sech}^{2}x$ and $\nu=1.2$, $\epsilon=10^{-2}$ at
$t=1$ and the corresponding asymptotic solution in terms of Hopf
and one-phase solution. The lower figure shows the same
difference, which is replaced close to the leading edge of the
Whitham zone by the difference between CH solution and multiscales
solution (shown in red where the error is smaller than the one
shown above).} \label{ch1e4match}
\end{figure}

\begin{figure}[!htb]
\centering \epsfig{figure=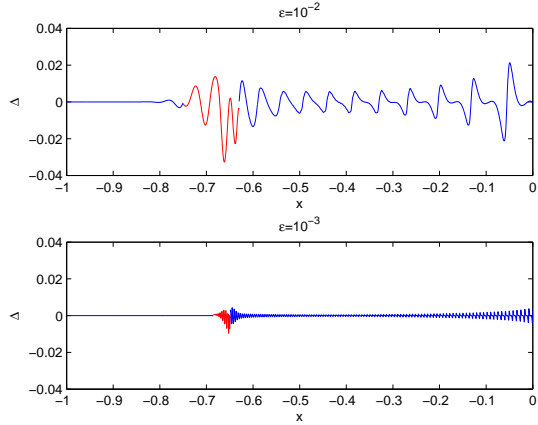, width=8cm} \caption{The
difference between the numerical solution to the CH equation for
the initial datum $u_{0}=-\mbox{sech}^{2}x$ and $\nu=1.2$ at $t=1$
and the corresponding asymptotic solution in terms of Hopf and
one-phase solution in blue and CH and multiscales solution in red,
where the latter error is smaller than the former, for two values
of $\epsilon$.} \label{chmatchp2}
\end{figure}

As can be already seen from Fig.~\ref{ch3in1}, the multiscales
solution gives a better asymptotic description of CH near the
leading edge of the Whitham zone than the Hopf and the one--phase
solution. This is even more obvious in Fig.~\ref{ch1e42d} where
the difference between CH and the asymptotic solutions is shown.
This suggests to identify the regions where each of the asymptotic
solutions gives a better description of CH than the other. The
results of this analysis can be seen in Fig.~\ref{ch1e4match}.
This matching procedure clearly improves the CH description near
the leading edge.
In Fig.~\ref{chmatchp2} we see the difference between this matched
asymptotic solution and the CH solution for two values of
$\epsilon$. Visibly the zone, where the solutions are matched,
decreases with $\epsilon$.

\begin{figure}[!htb]
\centering \epsfig{figure=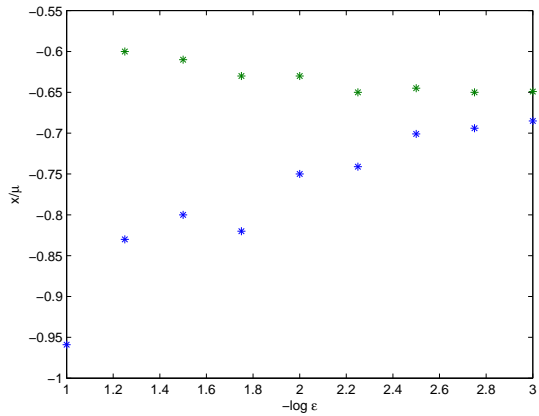, width=8cm} \caption{The edges
of the zone where the multiscales solutions provides a better
asymptotic description of CH than the Hopf or the one--phase
solution for the initial datum $u_{0}=-\mbox{sech}^{2}x$ and
$\nu=1.2$ at $t=1$ in dependence of $\epsilon$.} \label{width}
\end{figure}

There is a certain ambiguity in the precise definition of this
matching zone due to the oscillatory character of the solutions.
Because of the much lower number of oscillations than in KdV, the
statistics is considerably worse in the CH case than in the KdV
case. The limits of the matching zone for several values of
$\epsilon$ can be seen in Fig.~\ref{width}. Due to the lower
number of oscillations in the Hopf region, the matching zone
extends much further into this region than in the Whitham region.
The width of this zone scales like $\epsilon^{a}$ with $a=0.51$
and standard deviation $\sigma_{a}=0.06$ and correlation
coefficient $r=0.99$.

\section*{Acknowledgments}
This work has been supported by the MISGAM program of the European
Science Foundation and the project FroM-PDE funded by the European
Research Council through the Advanced Investigator Grant Scheme. CK
thanks for financial support by the Conseil R\'egional de Bourgogne
via the FABER scheme.


\end{document}